\documentclass[12pt,letterpaper]{article}

\usepackage[utf8]{inputenc}
\usepackage[T1]{fontenc}
\usepackage{lmodern}
\usepackage[english]{babel}
\usepackage[margin=1.25in]{geometry}
\usepackage{microtype}
\usepackage{setspace}
\usepackage{amsmath,amsthm,amssymb,mathtools}
\usepackage{booktabs,graphicx,enumitem,array,longtable,xcolor,tikz}
\usetikzlibrary{arrows.meta,positioning}
\usepackage[font=small,labelfont=bf]{caption}
\usepackage[authoryear,round]{natbib}
\usepackage[hidelinks,bookmarksnumbered=true]{hyperref}
\hypersetup{
  pdftitle={Correlation Probabilistic Sophistication},
  pdfauthor={Yi-Chun Chen; Soo Hong Chew; Xinhan Zhang}
}

\setlist[enumerate,1]{label=(\roman*)}

\theoremstyle{plain}
\newtheorem{thm}{Theorem}
\newtheorem{prop}{Proposition}
\newtheorem{lem}{Lemma}

\theoremstyle{definition}
\newtheorem{defi}{Definition}
\newtheorem{axm}{Axiom}
\theoremstyle{remark}

\newcommand{\Pf}{\mathcal P_f}
\newcommand{\Jf}{\mathcal J_f}

\newcommand{\T}{\mathsf T}
\newcommand{\R}{\mathbb R}
\newcommand{\E}{\mathbb E}

\newcommand{\Acts}{\mathcal A}
\newcommand{\Events}{\mathcal E}
\newcommand{\ind}{\mathbf 1}
\newcommand{\J}{\mathcal J}

\newcommand{\indif}{\widetilde\Pi}

\title{Correlation Probabilistic Sophistication}
\author{
  Yi-Chun Chen\thanks{Department of Economics and Risk Management Institute,
    National University of Singapore.
    Email: \href{mailto:yichun@nus.edu.sg}{\texttt{yichun@nus.edu.sg}}.}
  \and
  Soo Hong Chew\thanks{Center for Intelligence Economic Science,
    Southwestern University of Finance and Economics; and Department of Economics,
    National University of Singapore.
    Email: \href{mailto:chew.soohong@gmail.com}{\texttt{chew.soohong@gmail.com}}.}
  \and
  Xinhan Zhang\thanks{Center for Intelligence Economic Science,
    Southwestern University of Finance and Economics.
    Email: \href{mailto:zhangxinhan@swufe.edu.cn}{\texttt{zhangxinhan@swufe.edu.cn}}.}
}
\date{}

\begin{document}
\hypersetup{pageanchor=false}
\begin{titlepage}
\begin{singlespace}
\maketitle
\end{singlespace}
\thispagestyle{empty}
\vspace{-1em}
\begin{abstract}
\begin{singlespace}
\noindent
We study binary choice under uncertainty when the comparison between two acts can depend on the joint distribution of their realized outcomes. This motivates \emph{correlation probabilistic sophistication} (CPS), a subjective reduction principle that extends both probabilistic sophistication and Fishburn's nontransitive subjective expected utility. We show how simultaneous event exchangeability elicits a unique solvable finitely additive subjective probability and reduces binary comparisons of acts to induced joint laws. Under convex range, the induced domain coincides with the full finite joint-law domain, allowing the transfer of correlation-utility representations from risk and, under completeness and Strong Independence, correlation expected utility with a skew-symmetric comparison kernel. Under completeness and transitivity, CPS collapses to ordinary probabilistic sophistication. We further extend the framework to multiple small worlds, allowing correlation sensitivity to vary with the source of uncertainty.
\end{singlespace}
\end{abstract}
\vspace{0.5em}
\begin{quote}
\small\singlespacing
\noindent\textbf{Keywords:} correlation preference, probabilistic sophistication, event exchangeability, correlation expected utility, small worlds, nontransitive subjective expected utility.
\end{quote}
\vfill
\end{titlepage}
\setcounter{page}{1}
\hypersetup{pageanchor=true}

\section{Introduction}\label{sec:intro}

Savage's \citeyearpar{savage1954foundations} subjective expected utility (SEU) theory reduces choice under uncertainty to choice over induced lotteries: acts are evaluated by integrating utility with respect to a subjective probability on states. Two major post-Savage directions have loosened that reduction in different ways. One direction, developed by \citet{machina1992more}, retains completeness and transitivity but relaxes Savage's Postulate 2, yielding \emph{probabilistic sophistication} (PS): only the marginal lottery induced by an act matters, but the ranking over lotteries may be non-expected-utility. A second direction, developed by \citet{fishburn1989non}, keeps a sure-thing type separability while allowing nontransitivity; in modern language, this direction is correlation sensitive, because the comparison between two acts may depend on how their outcomes are paired state by state.

The present paper develops a framework that places these two directions in a common correlation-sensitive perspective. Our primitive object is not merely a pair of marginal lotteries, but the \emph{joint distribution} of the realized and forgone outcomes generated by a pair of acts under a subjective probability. A decision maker may care not only about the marginal distribution of each act, but also about whether favorable and unfavorable outcomes are aligned in the same states, whether gains are compared directly with nearby alternatives, or whether the pair generates particularly salient state-by-state contrasts. Such considerations arise naturally in regret theory \citep{bell1982regret,loomes1982regret}, salience theory \citep{bordalo2012salience}, and Fishburn's distribution-based analysis of binary choice \citep{fishburn1988nonlin}.

To see the basic issue, consider two acts that induce the same marginal distributions as two risky prospects, but are implemented on a common source of uncertainty. When the good outcome of one act systematically occurs in states where the other act performs poorly, the binary comparison can look very different from a comparison in which the same marginal distributions are generated independently. Under risk, this is precisely the distinction between independent implementation and a one-source, state-by-state comparison. Under uncertainty, the same phenomenon concerns how a subjective state space couples the two acts. Our goal is to formulate a subjective reduction principle that recognizes such coupling as behaviorally relevant, while preserving the spirit of probabilistic sophistication.

We call this principle \emph{correlation probabilistic sophistication} (CPS). Fix a finitely additive probability $\mu$ on the event algebra. Each ordered pair of acts $(f,g)$ induces a finite joint law $\pi_\mu[f,g]$ on realized-forgone outcome pairs. CPS requires that any two pairs of acts generating the same joint law yield the same binary comparison. Thus the joint law is the consequential object. Marginal-lottery reduction and ordinary probabilistic sophistication are recovered when the induced preference is insensitive to correlation and depends only on the marginals of the joint law. Fishburn's nontransitive subjective expected utility (SEU) is recovered on the opposite side through conditional separability and aggregation, while permitting correlation-sensitive, possibly nontransitive comparison.

\begin{figure}[h]
\centering
\begin{tikzpicture}[
  >=Stealth, line width=1.05pt,
  every node/.style={align=center,inner sep=3pt},
  model/.style={font=\large},
  edge label/.style={font=\footnotesize,inner sep=1pt}
]
  \node[model] (top) at (0,2.75) {\textbf{SEU}\\\small{S54}};
  \node[model] (left) at (-5,0) {\textbf{CSEU}\\\small{F89}};
  \node[model] (right) at (5,0) {\textbf{PS}\\\small{MS92}};
  \node[model] (bot) at (0,-2.75) {\textbf{CPS}\\\small{CCZ26}};
  \draw[-{Stealth[length=2.6mm,width=1.8mm]}] (top) -- (left)
    node[edge label,midway,sloped,above=2pt] {non-transitive}
    node[edge label,midway,sloped,below=2pt] {correlation-sensitive};
  \draw[-{Stealth[length=2.6mm,width=1.8mm]}] (top) -- (right)
    node[edge label,midway,sloped,above=2pt] {non-P2};
  \draw[-{Stealth[length=2.6mm,width=1.8mm]}] (left) -- (bot)
    node[edge label,midway,sloped,above=2pt] {non-P2};
  \draw[-{Stealth[length=2.6mm,width=1.8mm]}] (right) -- (bot)
    node[edge label,midway,sloped,above=2pt] {correlation-sensitive}
    node[edge label,midway,sloped,below=2pt] {non-transitive};
\end{tikzpicture}
\caption{A diamond of subjective models. CSEU denotes the correlation-sensitive subjective expected-utility form of Fishburn's model.}
\label{fig:diamond}
\end{figure}
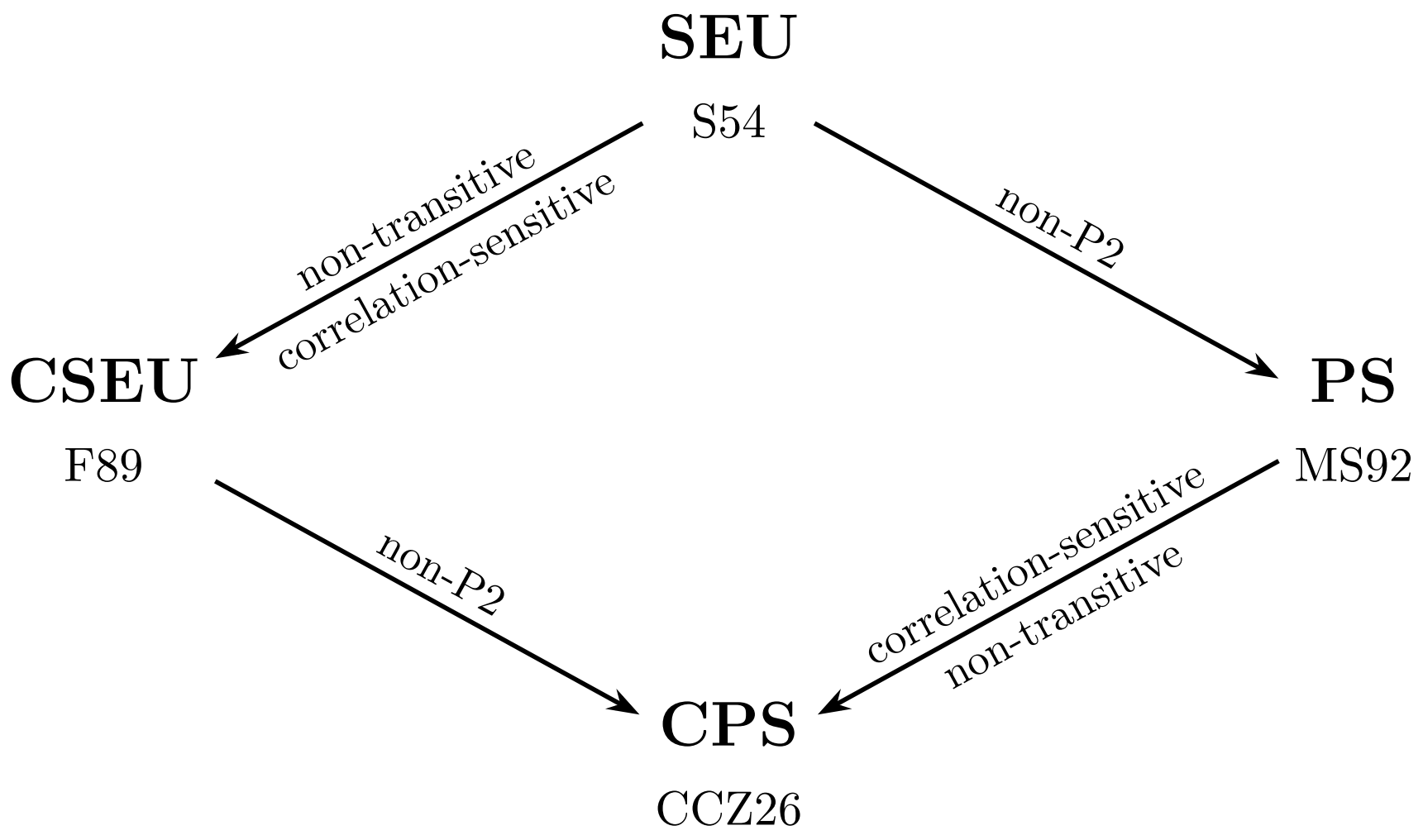

Figure~\ref{fig:diamond} summarizes this organizing role. SEU sits at the top as the fully transitive and correlation-insensitive benchmark. Moving left, Fishburn's nontransitive subjective expected utility preserves a form of statewise separability while allowing correlation sensitivity. Moving right, Machina--Schmeidler PS preserves transitivity while abandoning Savage's P2. Our contribution lies at the bottom: CPS contains both branches. CPS allows departures from Fishburn's conditional restrictions on the left and from transitivity and correlation insensitivity on the right. The diagram summarizes the principal modeling distinctions, while Appendix~\ref{app:fishburn} makes the left-hand branch precise: conditional aggregation, together with P2 and the stated regularity conditions, corresponds to Strong Independence \citep{lanzani2022correlation,chenChewZhang2026bang} on joint laws.

Our main technical device is a subjective extension of event exchangeability. We introduce \emph{simultaneous event exchangeability}: two disjoint events are exchangeable when swapping the realized outcomes on those two events in both acts, holding all background consequences fixed, never changes the binary comparison. This condition is the correlation-sensitive analogue of equal likelihood. By applying the event-exchangeability theorem of \citet{chew2006event} to an auxiliary ordering of pair-valued acts, we obtain a unique solvable, finitely additive probability that agrees with the induced comparative likelihood relation. The same argument establishes CPS: once probabilities are elicited in this way, binary comparisons depend only on the induced joint laws.

The theorem has two important consequences. First, it makes precise the link between subjective uncertainty and the risk-domain theory of correlation preference. If the elicited probability has \emph{convex range}, then every finite joint distribution over outcomes is realizable by some pair of acts, so the induced domain of subjective comparisons is exactly the full finite joint-law domain. As a result, the representation theorems developed for risk transfer directly: continuous correlation preference yields a correlation-utility representation \citep{chenChewZhang2026correlation}, and completeness together with Strong Independence yields subjective correlation expected utility (CEU), with a skew-symmetric comparison kernel. Appendix~\ref{app:fishburn} identifies the subjective restriction behind this CEU representation. With a sure-thing principle making conditional comparisons independent of common outside consequences, Fishburn's conditional aggregation axiom is equivalent to Strong Independence on the induced joint laws. This separates probability elicitation from the additional restriction that makes statewise comparisons additive. Whereas the companion paper by \citet{chenChewZhang2026bang} studies the risk-domain implications of correlation betweenness, the present paper develops the subjective foundation and the conditional restrictions on acts.

Second, CPS sharpens the role of transitivity. A central result of \citet{bikhchandani2011transitive} shows that transitivity can collapse certain regret-based comparisons back to probabilistic sophistication. We obtain a correlation-preference counterpart. Under completeness, transitivity, convex range, and closedness of the induced joint-law preference, every same-marginal joint law becomes indifferent; hence the induced comparison depends only on marginals, and CPS reduces to ordinary transitive probabilistic sophistication. In this sense, transitivity eliminates correlation sensitivity in the one-source Savage setting.

This reduction is informative precisely because correlation sensitivity is often source dependent. A state-by-state comparison generated within a familiar source of uncertainty may be evaluated differently from a comparison that combines uncertainties across distinct sources. We therefore extend CPS to the small-worlds approach of \citet{chew2008small}, allowing each source of uncertainty to have its own subjective probability and its own degree of correlation sensitivity. This source-dependent version offers a direct way to interpret phenomena such as Epstein and Halevy's \citeyearpar{epstein2019ambiguous} incremental aversion to correlation ambiguity.

The paper is organized as follows. Section~\ref{sec:setup} introduces the Savagean setup, defines CPS, and derives subjective probability from simultaneous event exchangeability. Section~\ref{sec:subjective} shows how CPS reduces uncertainty to the joint-law domain and yields subjective correlation-utility and subjective CEU representations. Section~\ref{sec:tps} proves that complete and transitive CPS collapses to ordinary probabilistic sophistication. Section~\ref{sec:smallworlds} extends the framework to multiple small worlds and source-dependent correlation sensitivity. Appendix~\ref{app:cps} collects the proofs and supporting results for these sections, including the small-worlds analysis. Appendix~\ref{app:fishburn} details the side-by-side comparison of Section \ref{sec:subjective} with \citet{fishburn1989non} through Conditional Aggregation and proves the equivalence with Strong Independence.

\section{CPS and Subjective Probability}\label{sec:setup}

\subsection{Savagean Acts and Joint-Distribution Reduction}

Let $S$ and $X$ be the nonempty sets of states and outcomes, and $\Events$ an algebra of events in $S$. The set of acts $\Acts$ consists of all $\Events$-measurable mappings from $S$ to a finite subset of $X$. Write $f\succeq g$ if the decision maker weakly prefers $f$ over $g$. We maintain nontriviality: there exist $f,g,h,k$ with $f\succeq g$ and $h\not\succeq k$. Define the strict preference $f\succ g$ if $f\succeq g$ and $g\not\succeq f$, and $f\sim g$ if both weak comparisons hold. We do not impose completeness and transitivity unless explicitly stated. Given a finitely additive probability $\mu$ on $\Events$, a pair of acts induces the joint law
\[
\pi_\mu[f,g](x,y):=\mu(\{s:f(s)=x,\ g(s)=y\}).
\]
Finite additivity is sufficient because the pair has finite range. The marginals of $\pi_\mu[f,g]$ are the distributions of $f$ and $g$; reversing the acts transposes the joint law.

\begin{defi}[Correlation Probabilistic Sophistication]\label{def:cps}
The decision maker exhibits \emph{correlation probabilistic sophistication} (CPS) with respect to $\mu$ if any two pairs of acts inducing the same joint distribution under $\mu$ give the same preference:
\[
\pi_\mu[f,g]=\pi_\mu[f',g']
\quad\Longrightarrow\quad
f\succeq g \text{ if and only if } f'\succeq g'.
\]
\end{defi}

CPS is a joint-distribution reduction principle. It says that the consequential content of a binary comparison is exhausted by the joint law of realized and forgone outcomes. Distinct implementations of the same two marginals may therefore be behaviorally different, but two implementations with the same joint law are not.

This principle sits between the one-coin and two-coin viewpoints emphasized by \citet{fishburn1988nonlin}. If the same pair of marginal lotteries is generated independently and under a common source of uncertainty, the resulting joint laws differ and CPS allows the decision maker to distinguish them. At the same time, CPS imposes a reduction: once the joint law is fixed, all further details of the state space become irrelevant for the comparison.

\subsection{Event Exchangeability and Subjective Probability}\label{subsec:exchangeability}

Write $f_Eh$ for an act coinciding with $f$ on $E$ and with $h$ outside $E$; analogous notation specifies several disjoint events. A constant act is identified with its outcome. Let $H:S\to X^2$ denote a finite-valued pair act, equivalently $H=(f,g)$. Write $I(H)=\ind_{\{f\succeq g\}}$ for its preference indicator. An event $E$ is \emph{null} if changing a pair act arbitrarily on $E$ leaves $I$ unchanged: $I(H_EK)=I(H'_EK)$ for all pair acts $H,H',K$. We offer the following definition as the correlation counterpart to event exchangeability in \citet{chew2006event}.

\begin{defi}[Simultaneous Event Exchangeability]\label{def:exchange}
Disjoint events $E,E'\in\Events$ are exchangeable, written $E\approx E'$, if
\[
 x_E x'_{E'}f\succeq y_E y'_{E'}g
 \quad\Longleftrightarrow\quad
 x'_E x_{E'}f\succeq y'_E y_{E'}g
\]
for every $x,x',y,y'\in X$ and $f,g\in\Acts$. For arbitrary events, write $E\approx E'$ when $E\setminus E'\approx E'\setminus E$.
\end{defi}

If $E\approx E'$, then simultaneously swapping the payoffs assigned to $E$ and $E'$ in the two acts never affects the comparison. This is naturally interpreted as a notion of equal likelihood. It yields the following comparative-likelihood relation: define $E\succeq_C E'$ if and only if some event $e\subseteq E\setminus E'$ satisfies $e\approx E'\setminus E$.

We apply the three conditions of \citet{chew2006event}: completeness of comparative likelihood, the Event Archimedean Property, and Event Nonsatiation.

\begin{axm}[Event Completeness, C]\label{ax:event-c}
For every disjoint $E,E'\in\Events$, either $E\succeq_C E'$ or $E'\succeq_C E$.
\end{axm}

\begin{axm}[Event Archimedean Property, A]\label{ax:event-a}
There is no infinite sequence $(e_n)_{n\geq0}$ of pairwise disjoint, nonnull events with $e_n\approx e_{n+1}$ for every $n$.
\end{axm}

\begin{axm}[Event Nonsatiation, N]\label{ax:event-n}
For pairwise disjoint $E,E',B\in\Events$, if $E\approx E'$ and $B$ is nonnull, then there is no $e\in\Events$, $e\subseteq E'$, with $e\approx E\cup B$.
\end{axm}

Event Completeness makes comparative likelihood complete. The Archimedean property prevents indefinitely many disjoint copies of a nonnull event. Nonsatiation prevents a copy of an event enlarged by a nonnull part from fitting inside an event of the original size.

A probability \emph{agrees} with $\succeq_C$ if $E\succeq_C E'$ exactly when $\mu(E)\geq\mu(E')$. It is \emph{solvable} if $\mu(E)\geq\mu(E')$ implies that some measurable $e\subseteq E$ has $\mu(e)=\mu(E')$.

\begin{thm}[Subjective Probability and CPS]\label{thm:cps}
The preference satisfies axioms C, A, and N if and only if there exists a unique solvable finitely additive agreeing probability $\mu$ for $\succeq_C$. In that case, $E\approx E'$ if and only if $\mu(E)=\mu(E')$, null events are exactly the zero-probability events, and $\succeq$ is CPS with respect to $\mu$.
\end{thm}

The proof applies the exchangeability theorem of \citet{chew2006event} to an auxiliary binary ordering of pair-valued acts. Order the finite-valued $X^2$-valued acts by the binary number $I(H)$. This auxiliary relation is a nondegenerate weak order; its exchangeable events and null events are exactly those defined above. Equality of the induced lotteries of the pair-valued acts is equality of the induced joint laws $\pi_\mu[f,g]$, giving CPS. Appendix~\ref{app:cps-proof} verifies these steps and provides a direct cycle-decomposition proof of the reduction from equal distributions of pair acts to equal preference indicators.

\section{From CPS to Correlation Expected Utility}\label{sec:subjective}

For the representation results, let $X$ be a metrizable outcome space with at least two elements. Write $\Pf(X)$ for the finitely supported Borel probability measures on $X$ and $\Jf(X):=\Pf(X^2)$ for their joint-law domain. For a finite $F\subset X$, let $\J(F):=\Pf(F^2)$. The marginals of $\pi$ are denoted $\pi_1$ and $\pi_2$, $\pi^\T$ is its transpose, and $\delta_{(x,y)}$ denotes unit mass on the outcome pair $(x,y)$. These domains carry their relative weak topology, tested against bounded continuous functions; on $\J(F)$ this is the usual finite-dimensional topology. Only limits remaining in the finite-support domain are considered.

Before transferring representations to acts, consider the risk-domain benchmark. A joint law $\pi\in\Jf(X)$ specifies the outcomes of two lotteries drawn together, and a preference set $\Pi\subseteq\Jf(X)$ records the laws at which the first lottery is weakly preferred to the second. Following \citet{lanzani2022correlation} and \citet{chenChewZhang2026bang}, a \emph{correlation expected utility} (CEU) representation evaluates these comparisons by a skew-symmetric kernel $\phi:X^2\to\R$, satisfying $\phi(x,y)=-\phi(y,x)$, such that
\begin{equation}\label{eq:risk-ceu}
 \pi\in\Pi
 \quad\Longleftrightarrow\quad
 U_\phi(\pi):=\E_\pi[\phi(x,y)]
 =\sum_{x,y}\phi(x,y)\pi(x,y)\geq0.
\end{equation}
The kernel measures the contribution of receiving $x$ while foregoing $y$; the expectation is over outcome pairs, so the comparison may depend on their coupling. Skew symmetry implies $\phi(x,x)=0$ and $U_\phi(\pi^\T)=-U_\phi(\pi)$. Hence CEU is complete, and positive, zero, and negative values identify strict preference, indifference, and reverse strict preference. Under nontriviality, the kernel is nonzero and takes both signs.

Expected utility is nested by $\phi(x,y)=u(x)-u(y)$: then $U_\phi(\pi)=\E_{\pi_1}u-\E_{\pi_2}u$ depends only on the marginals. CPS connects the CEU benchmark to acts by evaluating $\pi_\mu[f,g]$.

Let $\J_\mu:=\{\pi_\mu[f,g]:f,g\in\Acts\}$ be the realizable joint laws. CPS makes the induced preference set
\[
 \Pi_\mu:=\{\pi_\mu[f,g]:f\succeq g\}\subseteq\J_\mu
\]
well defined: every realization of a given joint law yields the same binary comparison. Nontriviality makes $\Pi_\mu$ nonempty and proper. Reversing the acts gives $\pi_\mu[g,f]=\pi_\mu[f,g]^\T$. The induced strict preference, indifference, and complement sets are
\begin{equation*}
 \widehat\Pi_\mu:=\{\pi:\pi\in\Pi_\mu,\ \pi^\T\notin\Pi_\mu\},\quad \widetilde\Pi_\mu:=\{\pi:\pi\in\Pi_\mu,\ \pi^\T\in\Pi_\mu\},\quad \check\Pi_\mu:=\J_\mu\setminus\Pi_\mu.
\end{equation*}
Completeness of the act relation is equivalent to $\pi\notin\Pi_\mu\implies \pi^\T\in\Pi_\mu$.

The probability $\mu$ has \emph{convex range} if, for every $E\in\Events$ and $t\in[0,1]$, there is $e\subseteq E$ with $\mu(e)=t\mu(E)$. This strengthens solvability, which matches probabilities of existing events.

\begin{lem}[Realization on an Event]\label{lem:realization}
If $\mu$ is convex-ranged, then for every $E$ with $\mu(E)>0$ and every finite probability vector $(t_1,\ldots,t_n)$, there is a measurable partition $E_1,\ldots,E_n$ of $E$ with $\mu(E_i)=t_i\mu(E)$. Consequently every finite joint law can be realized conditionally on $E$, and $\J_\mu=\Jf(X)$.
\end{lem}

\begin{proof}
Select the first $n-1$ cells successively from the residual event, using convex range to give the $i$th cell measure $t_i\mu(E)$. Give the last cell all remaining states. If the residual event has zero probability at an intermediate step, the remaining target masses are zero and the residual states can be assigned to the last cell. Finite additivity verifies the prescribed masses. Assigning each cell its prescribed outcome pair realizes the law.
\end{proof}

A finite space of equally likely atoms has a solvable probability but realizes masses only on a finite grid. Length probability on the algebra of intervals with rational endpoints is also solvable, with a dense rational range rather than convex range. Appendix~\ref{app:realization} records these examples. Thus the probability-elicitation theorem need not by itself provide the richness required to implement every mixture.

Under convex range, choose $E$ with $\mu(E)=\alpha$ and realize $\pi$ conditionally on $E$ and $\rho$ conditionally on $E^c$. The full joint law is $\alpha\pi+(1-\alpha)\rho$. This gives a direct Savagean implementation of the mixtures used in the risk-domain theory of correlation preference.

\begin{axm}[Continuity]\label{ax:closed}
The induced preference set $\Pi_\mu$ is closed relative to $\Jf(X)$ in the weak topology.
\end{axm}

The following is the joint-law Strong Independence condition used for CEU. Its first clause preserves indifference; its second preserves strict preference when combined with weak preference; its third is the corresponding restriction on the complement. Under completeness, the third clause is the transpose counterpart of the second.

\begin{axm}[Strong Independence]\label{ax:strong}
For $\pi,\rho\in\Jf(X)$ and $\alpha\in(0,1)$, with $\pi_\alpha:=\alpha\pi+(1-\alpha)\rho$:
\begin{enumerate}[noitemsep,topsep=2pt]
\item $\pi,\rho\in\widetilde\Pi_\mu$ implies $\pi_\alpha\in\widetilde\Pi_\mu$;
\item $\pi\in\widehat\Pi_\mu$, $\rho\in\Pi_\mu$ implies $\pi_\alpha\in\widehat\Pi_\mu$;
\item $\pi\in\check\Pi_\mu\cup\widetilde\Pi_\mu$, $\rho\in\check\Pi_\mu$ implies $\pi_\alpha\in\check\Pi_\mu$.
\end{enumerate}
\end{axm}

\begin{prop}[Subjective Correlation Utility and Subjective CEU]\label{prop:subjective}
Suppose $\succeq$ is CPS with respect to a convex-ranged finitely additive probability $\mu$. Then $\J_\mu=\Jf(X)$. If $\Pi_\mu$ satisfies Continuity, there is a nonconstant continuous $U:\Jf(X)\to\R$ such that $f\succeq g$ if and only if $U(\pi_\mu[f,g])\geq0$.

Under Continuity, $\succeq$ is complete and satisfies Strong Independence if and only if there is a nonzero skew-symmetric kernel $\phi:X^2\to\R$ such that
\begin{equation}\label{eq:subjective-ceu}
 f\succeq g
 \quad\Longleftrightarrow\quad
 U^\mu(f,g):=\sum_{x,y}\phi(x,y)\,
 \mu\big(\{s:f(s)=x,\ g(s)=y\}\big)\geq0.
\end{equation}
The kernel is unique up to a positive factor for the fixed probability. When $\mu$ is supplied by Theorem~\ref{thm:cps}, it is unique as an agreeing probability.
\end{prop}

The continuous correlation utility represents the sign of a binary comparison. CEU adds linearity in joint laws and skew symmetry of the comparison kernel, as in \citet{lanzani2022correlation} and \citet{chenChewZhang2026bang}. These restrictions yield complete comparisons while permitting nontransitivity. Appendix~\ref{app:realization} derives the subjective CEU representation from the risk-domain characterization in \citet[Corollary~2, Section~3.5]{chenChewZhang2026bang}, with a common normalization across finite outcome sets.

The subjective CEU representation connects our framework to the nontransitive subjective expected utility model of \citet{fishburn1989non}: both evaluate comparisons of acts by taking the subjective expectation of a skew-symmetric function of their statewise consequences. Appendix~\ref{app:fishburn} develops a side-by-side comparison of the axiom systems and establishes the correspondence between Fishburn's conditional aggregation and Strong Independence under the stated common assumptions.

\section[Transitivity and Ordinary Probabilistic Sophistication]{Transitivity and Probabilistic Sophistication}\label{sec:tps}

Section~\ref{sec:subjective} specializes CPS through restrictions on the induced joint-law preference without requiring transitivity. We now examine the transitive branch of Figure~\ref{fig:diamond}, imposing transitivity without the conditional restrictions studied in Appendix~\ref{app:fishburn}. A CPS relation is \emph{correlation insensitive} if its induced preference depends only on the marginal laws. It is ordinary transitive probabilistic sophistication (TPS) when it can be represented by a complete transitive relation $\succeq_L$ over the realizable marginal lotteries: $f\succeq g$ if and only if $\mu\circ f^{-1}\succeq_L\mu\circ g^{-1}$.

The following proposition adapts the finite-permutation argument of \citet[Proposition~1]{bikhchandani2011transitive} in \emph{Transitive Regret}. Their result shows how transitivity can collapse certain regret-based comparisons to marginal-lottery reduction. Here the primitive object is broader: CPS takes the full joint distribution of realized and forgone outcomes as behaviorally relevant. The proposition shows that transitivity eliminates even this broader form of correlation sensitivity in the one-source Savage setting.

\begin{prop}[Transitive CPS]\label{prop:tps}
Suppose $\mu$ is convex-ranged, $\succeq$ is complete and transitive, and $\succeq$ is CPS with respect to $\mu$. If the induced $\Pi_\mu$ is relatively weakly closed, then it is correlation insensitive, and $\succeq$ exhibits TPS.
\end{prop}

The proof has three steps. First consider a rational joint law with equal marginals. By convex range and Lemma~\ref{lem:realization}, it can be realized on an equally likely finite partition. Equality of marginals means that the two acts assign the same multiset of outcomes to the cells, so one act is a permutation of the other. Repeating this permutation generates a finite cycle of acts in which every consecutive comparison has the same joint law. CPS therefore assigns the same orientation to each consecutive comparison, and completeness plus transitivity rule out strict preference around the cycle. Hence the original two acts must be indifferent.

Second, rational balanced joint laws are dense in every finite same-marginal set. Relative weak closedness then extends the indifference conclusion from rational balanced laws to all balanced laws. Finally, once any two acts with the same marginal law are indifferent, transitivity allows either side of any comparison to be replaced by another act with the same law. The comparison therefore depends only on the marginal lotteries, which yields a well-defined complete and transitive marginal relation. Appendix~\ref{app:tps-proof} gives the formal proof.

Proposition~\ref{prop:tps} identifies transitivity as the decisive force behind the collapse from CPS to PS. Correlation sensitivity is therefore compatible with subjective reduction only when one permits nontransitivity or incompleteness, or when one enlarges the framework beyond a single undifferentiated source of uncertainty. The next section pursues the latter route.

\section{Small Worlds and Source-Dependent CPS}\label{sec:smallworlds}

Under the Savagean perspective, lotteries are jointly determined by uncertainties arising from a single grand world. In this section, we explore the implications of having multiple sources of uncertainty by adapting the small-worlds approach of \citet{chew2008small}.

A small world is a specified subalgebra $\Events_k\subseteq\Events$ together with its domain $\Acts_k$ of finite-valued measurable acts. A bet belongs to the small world with respect to which it is measurable. Each source has a nontrivial relation $\succeq_k$ on $\Acts_k$. Building on Section~\ref{sec:setup}, we adapt event exchangeability to the small-worlds setting by restricting events, acts, and background comparisons to the source. This yields source-specific comparative likelihood and null events.

\begin{prop}[CPS within a Source]\label{prop:source}
For each source $k$, suppose Event Completeness, the Event Archimedean Property, and Event Nonsatiation hold on that source. Then there is a unique solvable finitely additive agreeing probability $\mu_k$ on $\Events_k$, and $\succeq_k$ is CPS with respect to $\mu_k$. If $\mu_k$ is convex-ranged, Proposition~\ref{prop:subjective} applies to the induced source-specific preference set; in particular, Continuity, Completeness, and Strong Independence yield a source-specific CEU representation.
\end{prop}

Depending on the small world associated with the binary choice situation, the decision maker can possess varying degrees of correlation sensitivity, ranging from being transitive and fully correlation insensitive to being sensitive even on same-marginal comparisons. In a CEU specification, the source dependence is expressed by source-specific skew-symmetric kernels $\phi_k$ and comparison values $U^k(\pi)=\E_\pi\phi_k$. For comparisons across small worlds, the decision maker can rely on certainty equivalents of the bets assessed within their respective sources.

Compatibility across sources can be imposed in different ways. If the $\succeq_k$ are restrictions of one unlabelled relation, then $f\succeq_k g$ agrees with $f\succeq_\ell g$ whenever $f,g\in\Acts_k\cap\Acts_\ell$. Common beliefs on overlapping events require $\mu_k(E)=\mu_\ell(E)$ for $E\in\Events_k\cap\Events_\ell$. Alternatively, source labels can be treated as part of the comparison context, so that differently labelled choice problems are distinct primitives.

A useful illustration comes from \citet{epstein2019ambiguous}'s discussion of correlation ambiguity in the two-urn Ellsberg environment. There are two urns numbered 1 and 2, each containing two balls that are either red or black, with unknown compositions. One ball is drawn from each urn simultaneously. Let $R_1$ and $B_1$ denote the color events in urn 1, and let $\mathrm{Same}$ and $\mathrm{Diff}$ denote matching and different colors across the two draws. Consider the choice pattern
\[
R_1\succ\mathrm{Same}
\qquad\text{and}\qquad
B_1\succ\mathrm{Diff}.
\]
As observed by \citet{epstein2019ambiguous}, this pattern is inconsistent with a single probability satisfying $\mu(R_1)>\mu(\mathrm{Same})$ and $\mu(B_1)>\mu(\mathrm{Diff})$, because the two pairs of events partition the same state space.

Small-worlds CPS offers a direct interpretation. View $R_1$ and $B_1$ as bets within the single-urn source, and $\mathrm{Same}$ and $\mathrm{Diff}$ as bets within the mixed source. In each source, the relevant events are exchangeable, so the source-specific probabilities can be symmetric. Normalize the prize to one. In the first source, let the two complementary color events each have probability one half and use $u_1(z)=z$ on $[0,1]$. In the second source, let $\mathrm{Same}$ and $\mathrm{Diff}$ each have probability one half and use $u_2(z)=\sqrt z$. The within-source certainty equivalents are then $1/2$ and $1/4$, respectively. Comparing these certainty equivalents across sources yields $R_1\succ\mathrm{Same}$ and $B_1\succ\mathrm{Diff}$. Thus the decision maker is EU in both small worlds, but more ambiguity averse in the mixed source.

For a source relation, define the certainty-equivalent set $C_k(f):=\{c\in X:f\sim_k c\}$. In subjective CEU, skew symmetry gives
\[
 C_k(f)=\left\{c\in X:\sum_x\phi_k(x,c)\,
       \mu_k\big(\{s:f(s)=x\}\big)=0\right\}.
\]
If $X=[a,b]$ and the CEU kernel $\phi_k$ is continuous and strictly increasing in its first argument, every act in source $k$ has a unique monetary certainty equivalent. Appendix~\ref{app:ce} proves this claim. A rule comparing source-specific certainty equivalents then specifies cross-source choice.

The small-worlds formulation therefore preserves the main insight of CPS while escaping the collapse result of Proposition~\ref{prop:tps}. Correlation sensitivity need not disappear globally once different sources of uncertainty are kept distinct. This makes the framework suitable for modeling context-dependent or source-dependent comparison effects under uncertainty.

\section{Conclusion}\label{sec:conclusion}

This paper develops correlation probabilistic sophistication as a subjective foundation for correlation-sensitive binary choice under uncertainty. CPS reduces comparisons of acts to induced joint distributions of realized and forgone outcomes, thereby unifying the two classic post-Savage branches represented by Fishburn's nontransitive subjective expected utility and Machina--Schmeidler probabilistic sophistication. Simultaneous event exchangeability elicits a unique agreeing subjective probability and yields the reduction to joint laws. Under convex range, Continuity yields a continuous correlation-utility representation, and adding Completeness and Strong Independence yields subjective CEU. Conditional aggregation, together with the Sure-Thing Principle and the stated regularity conditions, characterizes Strong Independence and the subjective CEU representation. Under completeness and transitivity, CPS collapses to ordinary probabilistic sophistication. Extending the framework to small worlds restores room for source-dependent correlation sensitivity and provides a natural interpretation of incremental aversion to correlation ambiguity.

\begingroup
\singlespacing
\small
\setlength{\bibsep}{0.6\baselineskip}
\bibliographystyle{plainnat}
\bibliography{ref}
\endgroup

\appendix
\setcounter{table}{0}
\renewcommand{\thetable}{\thesection.\arabic{table}}

\section{CPS and Subjective Representations}\label{app:cps}

\subsection{Probability Elicitation and the CPS Proof}\label{app:cps-proof}

We use \citet[Theorem 1]{chew2006event}. On all simple acts with values in an arbitrary payoff set, a nondegenerate weak order induces exchangeability by indifference under swaps of constant payoffs on disjoint events. For that exchangeability and its associated null events and comparative likelihood, C, A, and N are equivalent to a unique solvable finitely additive agreeing probability. Equal probabilities characterize exchangeability.

\begin{proof}[Proof of Theorem \ref{thm:cps}]
Let $\mathcal H$ be all finite-valued $\Events$-measurable maps from $S$ to $X^2$. Every $H\in\mathcal H$ is uniquely a pair $(f,g)$ with $f,g\in\Acts$, and every such pair is in $\mathcal H$. Define an auxiliary relation by $H\succeq^a K$ if and only if $I(H)\geq I(K)$, where $I((f,g))=\ind_{\{f\succeq g\}}$.
It is complete and transitive because it compares real numbers. Nontriviality of $\succeq$ gives a pair with auxiliary value one and a pair with value zero, so it is nondegenerate. Its indifference relation is equality of preference indicators.

For disjoint $E,E'$, swapping constant payoffs $z=(x,y)$ and $z'=(x',y')$ in an auxiliary act produces the two comparisons in the definition of Simultaneous Event Exchangeability. Consequently auxiliary event exchangeability coincides exactly with Definition \ref{def:exchange}, including its universal quantification over background pair acts. An event is null for the auxiliary relation if arbitrary changes of the auxiliary act on it leave its value unchanged. This is exactly the comparison-status definition of nullity used in Section \ref{sec:setup}. Comparative likelihood, being defined only from event inclusion and exchangeability, is the same for both constructions. Thus all three event axioms coincide.

The cited theorem applies to $(S,\Events,X^2,\mathcal H,\succeq^a)$. It gives the unique agreeing probability and solvability, as well as the equivalence between equal probabilities and exchangeability. The reverse implication follows by applying the reverse direction of that theorem to the same auxiliary ordering.

We give a direct verification of CPS from the resulting probability and exchanges. First, behavioral null events have measure zero: if $E$ is null, it is exchangeable with the empty event, and equality of probabilities gives $\mu(E)=0$. Conversely, if $\mu(E)=0$, every measurable subevent has probability zero and is exchangeable with the empty event. For two pair acts differing only on $E$, refine $E$ into finitely many cells on which both are constant. On each cell, exchange with the empty event permits replacement of its constant pair payoff. These finitely many replacements preserve the preference indicator. Thus $E$ is null.

Now let $H,K\in\mathcal H$ have the same distribution under $\mu$. Enumerate their joint range as $z_1,\ldots,z_m$ and write $E_{ij}=\{s:H(s)=z_i,\ K(s)=z_j\}$. The matrix $t_{ij}=\mu(E_{ij})$ has equal row and column sums. Its off-diagonal entries therefore constitute a balanced finite flow. If this flow is nonzero, it contains a directed cycle of positive edges: follow outgoing positive flow, which exists at every vertex with positive incoming flow, until a vertex repeats.

Choose such a cycle and let $t>0$ be its smallest edge mass. In every edge cell of the cycle, solvability gives a subevent of measure $t$, using a cell with the smallest mass as the probability to be matched. On these disjoint equal-probability subevents, the values of $H$ can be permuted cyclically to agree with $K$. A finite sequence of transpositions implements the cycle. Every transposition is a simultaneous exchange of constant pair payoffs on exchangeable events, with all other values held as background, and hence leaves $I$ unchanged.

Remove the chosen subevents from the residual cells. Balance is preserved and at least one positive edge mass disappears. Repeat. There are only finitely many initially positive off-diagonal edges, so the procedure terminates after finitely many cycles. Any residual disagreement is confined to a finite union of zero-probability cells and is null. Each step preserves the preference indicator, proving $I(H)=I(K)$. Taking $H=(f,g)$ and $K=(f',g')$ gives CPS.
\end{proof}

\subsection{Realizability and Transfer of the Risk Results}\label{app:realization}

We give the topological argument for continuous correlation utility and transfer the risk-domain CEU characterization to acts. The arguments use the induced set $\Pi_\mu\subseteq\Jf(X)$ and introduce no further restrictions on the state space.

\paragraph*{Topology and Separation.}
\begin{lem}[Metric on the Finite-Support Domain]\label{lem:metric}
Let $Y$ be metrizable and let $d\leq1$ be a compatible bounded metric. On $\Pf(Y)$, the bounded-Lipschitz metric
\[
 \beta(\nu,\eta)
 :=\sup_{\substack{\|h\|_\infty\leq1\\|h(z)-h(z')|\leq d(z,z')}}
       \left|\int h\,d\nu-\int h\,d\eta\right|
\]
induces the relative weak topology.
\end{lem}

\begin{proof}
The displayed expression is a metric: bounded Lipschitz functions separate finite signed atomic measures, as can be seen using disjoint tent functions around their finitely many support points. Symmetry and the triangle inequality follow directly from the supremum.

Fix $\eta=\sum_{i=1}^n a_i\delta_{z_i}$. For a small $\epsilon>0$, choose pairwise disjoint balls about the $z_i$ of radii at most $\epsilon$. Let $\eta_i$ be nonnegative Lipschitz tent functions, equal to one at $z_i$, zero outside the corresponding ball, and satisfying $\sum_i\eta_i\leq1$. For every function $h$ in the supremum defining $\beta$, $\left|\int h\,d\nu-\sum_i a_i h(z_i)\right|\leq\epsilon+2\sum_i\left|\int\eta_i\,d\nu-a_i\right|$.
Indeed, insert $\sum_i\eta_i h$ into the integral, use
$|h(z)-h(z_i)|\leq\epsilon$ on the support of $\eta_i$, and bound the remaining mass by $1-\sum_i\int\eta_i\,d\nu$. The right side is uniformly small on a weak neighborhood of $\eta$. Thus weak convergence to a finitely supported limit implies convergence in $\beta$.

Conversely, let $g\in C_b(Y)$. Choose the balls so that $|g(z)-g(z_i)|<\epsilon$ within the $i$th ball. Convergence in $\beta$ implies convergence of each tent-function integral, because a fixed bounded Lipschitz function is a positive scalar multiple of a function allowed in its definition. The same decomposition gives $\left|\int g\,d\nu-\int g\,d\eta\right|\leq\epsilon+2\|g\|_\infty\sum_i\left|\int\eta_i\,d\nu-a_i\right|$.
Hence every bounded-continuous test integral is continuous for $\beta$. Both arguments are neighborhood arguments and apply to nets as well as sequences.
\end{proof}

\begin{proof}[Proof of Proposition \ref{prop:subjective}]
Lemma~\ref{lem:realization} gives $\J_\mu=\Jf(X)$. CPS makes $\Pi_\mu$ a nonempty proper subset of this domain, and transposition transfers the two oriented comparisons. For the continuous utility claim, apply Lemma~\ref{lem:metric} to $Y=X^2$ and set
\[
 U(\pi):=-\inf_{\rho\in\Pi_\mu}\beta(\pi,\rho).
\]
Distance is continuous, and closedness makes its zero set exactly $\Pi_\mu$. Since $\Pi_\mu$ is nonempty and proper, $U$ is nonconstant and has the required sign representation.

For CEU, suppose Completeness and Strong Independence also hold. Choose $\pi^-\in\check\Pi_\mu$ and put $\pi^+=(\pi^-)^\T$. Completeness makes $\pi^+$ strict. Let $F_0$ contain every outcome appearing in these laws. On every finite $F\supseteq F_0$, the set $\Pi_\mu\cap\J(F)$ is nonempty, proper, and closed, and it inherits Completeness and Strong Independence. Relabeling $F$ as a finite subset of $\R$ if necessary, apply \citet[Corollary~2, Section~3.5]{chenChewZhang2026bang} to obtain a skew-symmetric CEU kernel $\phi_F$ representing that set.

On each such face, the kernel is unique up to a positive factor. Indeed, both signs occur in the relative interior of the joint-law simplex, so the boundary of its weak-preference halfspace determines the zero hyperplane of any representing linear functional. Two such functionals therefore are proportional, with a positive factor because their signs agree. Normalize $\phi_F$ so that $\sum_{x,y}\phi_F(x,y)\pi^+(x,y)=1$.

If $F_0\subseteq F\subseteq G$, the kernel on $G$ restricts to an exact representation on $F$ with the same normalization. Uniqueness on $F$ makes these restrictions identical. For two faces that are not nested, use their finite union. Consequently, for any $x,y\in X$, define $\phi(x,y)$ from any finite face containing $F_0\cup\{x,y\}$; the definition is independent of the chosen face. The resulting kernel is skew-symmetric, nonzero, and represents $\Pi_\mu$ on every finite joint law. Substitution of $\pi_\mu[f,g]$ yields \eqref{eq:subjective-ceu}. Skew symmetry gives $U^\mu(g,f)=-U^\mu(f,g)$, so strict preference and indifference correspond exactly to positive and zero values.

Any other CEU kernel for the fixed probability, normalized at $\pi^+$ in the same way, agrees with $\phi$ on every finite face by uniqueness on that outcome set. This proves global uniqueness up to a positive factor. Conversely, linearity in joint laws and skew symmetry verify Completeness and all three clauses of Strong Independence. Probability uniqueness, when elicited by the event theorem, follows from Theorem~\ref{thm:cps}.
\end{proof}

\paragraph*{Solvability versus Convex Range.}
On $n$ equally likely atoms, subevents match the probability of any smaller existing event, while their probabilities are multiples of $1/n$. For a nonatomic example, let $S=[0,1)$ and let $\Events$ be the algebra of finite unions of half-open intervals with rational endpoints. Length is finitely additive, nonatomic, and solvable: any smaller rational length can be cut from the finitely many component intervals using rational endpoints. Its range is rational. In both examples, $\J_\mu$ records the realizable probability masses; solvability alone does not make it the full joint-law domain.

\subsection{Transitivity Reduction}\label{app:tps-proof}

\begin{proof}[Proof of Proposition \ref{prop:tps}]
We first apply a finite permutation to equally likely cells, following \citet[Proposition~1]{bikhchandani2011transitive}. Closedness on the joint-law domain then gives the passage from rational to arbitrary finite laws.

First take a rational $\pi\in\J(F)$ with equal marginals and let $N$ be a common denominator for its masses. By Lemma~\ref{lem:realization}, realize it as $(f_0,f_1)$ on $N$ equally likely cells. Equality of marginals means that the vectors of values of $f_0$ and $f_1$ contain the same multiset. There is a permutation $\sigma$ of the cells with $f_1(i)=f_0(\sigma i)$.

For $k\geq0$, set $f_k(i)=f_0(\sigma^k i)$. For every $k$, the law of $(f_k,f_{k+1})$ equals $\pi$, since $\sigma^k$ permutes equal-probability cells. A positive power of the finite permutation is the identity, so this sequence returns to $f_0$. CPS makes all consecutive comparisons have the same two oriented preference indicators. If either orientation were strict, completeness and transitivity would imply a strict cycle, which is impossible. Thus $f_0\sim f_1$ and $\pi\in\indif_\mu$.

The set of balanced matrices $\{\pi\in\J(F):\pi_1=\pi_2\}$ is a polytope described by rational linear equalities and inequalities, so its rational points are dense. Equivalently, decompose a balanced flow into finitely many directed cycles and approximate the weights of their normalized cycle laws by rational weights summing to one. Hence any balanced $\pi\in\J(F)$ is the limit of rational balanced laws $\pi_n$. Both $\pi_n$ and $\pi_n^\T$ belong to $\Pi_\mu$. Relative weak closedness and continuity of transposition give $\pi,\pi^\T\in\Pi_\mu$.

Any two acts with the same marginal law therefore are indifferent. For acts $f,g,f',g'$ with matching respective marginal laws, we have $f\sim f'$ and $g\sim g'$. Transitivity implies $f\succeq g$ if and only if $f'\succeq g'$. This is correlation insensitivity. Define $\succeq_L$ by any realizations of the marginals. The equivalence just proved makes it well defined; completeness and transitivity transfer from $\succeq$. This establishes TPS.
\end{proof}

\subsection{Small Worlds and Certainty Equivalents}\label{app:smallworlds}\label{app:ce}

The sourcewise claims in Proposition~\ref{prop:source} follow by applying Theorem~\ref{thm:cps} and, under convex range, Proposition~\ref{prop:subjective} separately within each source. We record here the sufficient condition used in Section~\ref{sec:smallworlds} for a unique monetary certainty equivalent.

\begin{prop}[Unique Monetary Certainty Equivalent]\label{prop:ce}
Suppose $X=[a,b]$, and source $k$ admits a subjective CEU representation with a continuous kernel $\phi_k$ that is strictly increasing in its first argument. Then every act in source $k$ has a unique monetary certainty equivalent.
\end{prop}

\begin{proof}
Fix an act $f$ and let $p(x)=\mu_k(\{s:f(s)=x\})$ be its induced lottery. By the CEU representation and skew symmetry, $f\sim_k c$ if and only if
\[
 V(c):=\sum_x p(x)\phi_k(x,c)=0.
\]
Continuity of $\phi_k$ makes $V$ continuous. Since $\phi_k(x,a)\geq\phi_k(a,a)=0$ and $\phi_k(x,b)\leq\phi_k(b,b)=0$ for all $x\in[a,b]$, we have $V(a)\geq0$ and $V(b)\leq0$. Skew symmetry and strict increase in the first argument make $\phi_k(x,c)=-\phi_k(c,x)$ strictly decreasing in $c$, so $V$ is strictly decreasing. The intermediate value theorem gives existence, and strict monotonicity gives uniqueness.
\end{proof}

\section{Fishburn's Conditional Restriction}\label{app:fishburn}

\subsection{Conditional Aggregation and Strong Independence}\label{app:fishburn-aggregation}

The subjective CEU representation in Section~\ref{sec:subjective} also clarifies the connection to \citet{fishburn1989non}. His conditional aggregation axiom is the event-based counterpart of Strong Independence. To formulate it, first separate the comparison on an event from the common consequences outside it.

\begin{axm}[Sure-Thing Principle for Common Consequences]\label{ax:stp}
For every $f,g,h,k\in\Acts$ and $E\in\Events$,
\[
 f_Eh\succeq g_Eh
 \quad\Longleftrightarrow\quad
 f_Ek\succeq g_Ek.
\]
\end{axm}

Under completeness, this is the weak-preference formulation of P2. Define $f\succeq_E g$ if $f_Eh\succeq g_Eh$ for every common background $h$; by the axiom, one background suffices. Write $f\succ_E g$ when $f\succeq_Eg$ but $g\not\succeq_Ef$, and $f\sim_Eg$ when both weak comparisons hold. These conditional relations are complete whenever $\succeq$ is complete.

\begin{axm}[Conditional Aggregation]\label{ax:aggregation}
For disjoint events $E,D$,
\[
 f\succeq_Eg\ \text{ and }\ f\succeq_Dg
 \quad\Longrightarrow\quad
 f\succeq_{E\cup D}g.
\]
If either conditional comparison is strict, the comparison on $E\cup D$ is strict.
\end{axm}

This is the restriction corresponding to Fishburn's $S1^*$. It aggregates comparisons of the \emph{same two acts} across disjoint events; it does not impose transitivity across three acts. P2 makes the conditional comparisons independent of the common background, while Conditional Aggregation governs how those comparisons combine.

\begin{prop}[Conditional Aggregation and Strong Independence]\label{prop:fishburn}
Suppose $\succeq$ is complete, is CPS with respect to a convex-ranged $\mu$, and satisfies the Sure-Thing Principle for Common Consequences. If the induced $\Pi_\mu$ satisfies Continuity, then Conditional Aggregation holds if and only if $\Pi_\mu$ satisfies Strong Independence.
\end{prop}

Together with Proposition~\ref{prop:subjective}, the result yields the subjective CEU representation. Its interpretation is immediate from the conditional comparison value
\[
 U_E^\mu(f,g):=\sum_{x,y}\phi(x,y)\,
       \mu\big(E\cap\{s:f(s)=x,\ g(s)=y\}\big).
\]
The common background contributes zero because $\phi(c,c)=0$, and for disjoint events,
$U_{E\cup D}^\mu(f,g)=U_E^\mu(f,g)+U_D^\mu(f,g)$.
Thus favorable conditional comparisons aggregate, retaining strictness if one component is strict. Conversely, convex range lets conditional aggregation operate on arbitrary mixtures of joint laws. The nontrivial step is to show that embedding a comparison in an event of any positive probability preserves its weak, strict, or indifferent status. Appendix~\ref{app:fishburn-proof} proves this step and the equivalence.

\subsection{Proof of the Equivalence}\label{app:fishburn-proof}

\begin{proof}[Proof of Proposition \ref{prop:fishburn}]
Assume conditional aggregation. Fix a constant $c$ and write $D_t(\rho)=t\rho+(1-t)\delta_{(c,c)}$ for a finite joint law $\rho$ and $t\in(0,1]$. If a pair $(f,g)$ has conditional law $\rho$ on an event $E$ of probability $t$, then $f\succeq_E g$ exactly when $D_t(\rho)\in\Pi_\mu$: use the common constant background $c$, CPS, and the Sure-Thing Principle for Common Consequences. Conditional strictness and indifference have the same correspondence. We first show that $D_t(\rho)$ and $\rho$ have identical behavioral categories.

For $t=1/n$, partition $S$ into $n$ equal-probability events and realize conditional law $\rho$ on every event. Each cell has the same conditional category by CPS with the constant background. If $\rho\in\Pi_\mu$ but the row alternative were conditionally strictly worse on each cell, iterating conditional aggregation would imply reverse strictness globally, a contradiction. Thus each cell gives conditional weak preference for the row alternative. If $\rho\in\widehat\Pi_\mu$ but each cell gave indifference, aggregation in both orientations would give global indifference, again a contradiction. Reversing orientations covers $\rho\in\check\Pi_\mu$, and applying the argument in both orientations covers indifference. Therefore all categories agree at $t=1/n$.

Aggregating $k$ such cells gives the same conclusion at every rational $t=k/n\in(0,1]$. For arbitrary $t$, approximate it by rational $t_j$. If $\rho\in\Pi_\mu$, then $D_{t_j}(\rho)\in\Pi_\mu$, and closedness gives $D_t(\rho)\in\Pi_\mu$. Indifference follows by transposition. To preserve strictness for an arbitrary $t$, take a rational $s\in(0,t)$ and an event $E$ of probability $t$. Realize conditional law $\rho$ on $E$, subdividing every outcome-pair cell in the proportion $s/t$ to obtain an event $D\subset E$ of probability $s$ with the same conditional law on $D$ and $E\setminus D$. This subdivision is possible by convex range. If $\rho\in\widehat\Pi_\mu$, the row alternative is conditionally strictly preferred on $D$ by the rational case and weakly preferred on $E\setminus D$ by the arbitrary-probability case. Conditional aggregation gives strict preference on $E$. Reversing orientations covers $\check\Pi_\mu$. Thus all categories are preserved for every $t>0$.

Now realize $\pi$ on an event of probability $\alpha$ and $\rho$ on its complement. If both laws belong to $\Pi_\mu$, conditional aggregation places their mixture in $\Pi_\mu$. If one is strict, it makes the mixture strict. If both are indifferent, applying the argument to both orientations gives indifference. Finally, if $\pi\in\check\Pi_\mu$ and $\rho\in\check\Pi_\mu\cup\widetilde\Pi_\mu$, completeness makes the reversed first law strict and the reversed second law weakly preferred; their reversed mixture is strict. Thus the original mixture belongs to $\check\Pi_\mu$. These are the three clauses of Strong Independence.

Conversely, Proposition~\ref{prop:subjective} gives a skew kernel for the induced preference. With a common background outside $E$, its total utility value is
$\sum_{x,y}\phi(x,y)\mu(E\cap\{f=x,g=y\})$, because the common background contributes only zero diagonal terms. This utility value is additive across disjoint events. Nonnegative conditional utility values therefore aggregate to a nonnegative utility value, and strict positivity on one event makes the sum positive. This is conditional aggregation. A zero-probability event contributes zero and can never supply the strict premise.
\end{proof}

In subjective CEU the representation is an expectation of a skew-symmetric function of the two consequences, as in \citet{fishburn1989non}. His probability elicitation and binary-outcome ordering conditions provide a different route to that form.

%\clearpage
\subsection{Comparison with Fishburn's Axioms}\label{app:fishburn-comparison}

The result makes precise the Fishburn branch in Figure~\ref{fig:diamond}: the relevant specialization uses P2 \emph{and} Conditional Aggregation, not P2 alone. Table~\ref{tab:fishburn} compares the roles of $P1^*$, $P2$--$P5$, $P6^*$, and $S1^*$ with those of the present formulation. The equivalence in Proposition~\ref{prop:fishburn} concerns $S1^*$ and Strong Independence under its common hypotheses; it does not identify the two complete axiom systems.

\begingroup
\singlespacing
\small
\setlength{\tabcolsep}{4pt}
\renewcommand{\arraystretch}{1.00}
\begin{longtable}{@{}>{\raggedright\arraybackslash}p{.12\textwidth}>{\raggedright\arraybackslash}p{.40\textwidth}>{\raggedright\arraybackslash}p{.42\textwidth}@{}}
\caption{Fishburn's conditions and their relation to CPS.}\label{tab:fishburn}\\
\toprule
Condition & Role in Fishburn's formulation & Relation to the present results\\
\midrule
\endfirsthead
\toprule
Condition & Role in Fishburn's formulation & Relation to the present results\\
\midrule
\endhead
$P1^*$ & Asymmetric strict preference and a weak ordering on each two-outcome act domain. & Completeness is maintained in Proposition \ref{prop:fishburn}; the probability theorem permits incompleteness of the original relation.\\[3pt]
$P2$ & Independence of a strict comparison from common consequences outside an event. & Under completeness, its weak formulation is the Sure-Thing Principle for Common Consequences, imposed in the conditional equivalence.\\[3pt]
$P3$ & State-independent comparison of constant outcomes on nonnull events. & In subjective CEU, multiplication by $\mu(E)>0$ preserves the sign of $\phi(x,y)$.\\[3pt]
$P4$ & Comparative likelihood inferred from binary bets is independent of the chosen ordered prizes. & Replaced in the probability argument by simultaneous event exchangeability and C, A, N.\\[3pt]
$P5$ & Some strictly ordered constant outcomes. & Strengthens the nontriviality of the act relation maintained in Section~\ref{sec:setup}.\\[3pt]
$P6^*$ & Finite partitions into sufficiently small events preserve a strict comparison under specified local payoff changes. & Relative weak closedness and convex range provide continuity and richness in Proposition \ref{prop:fishburn}.\\[3pt]
$S1^*$ & Weak conditional comparisons on disjoint events aggregate, with strictness if one component is strict. & Proposition \ref{prop:fishburn} proves its equivalence to Strong Independence under the listed common conditions.\\
\bottomrule
\end{longtable}
\endgroup

\end{document}